\documentclass[aps,pra,twocolumn,floatfix,showpacs,superscriptaddress]{revtex4}

\usepackage{graphicx}
\usepackage{epsfig}
\usepackage{bbm}
\usepackage[utf8]{inputenc}
\usepackage{color}
\usepackage{amsmath,amssymb}
\usepackage{natbib}
\usepackage[normalem]{ulem}

\begin{document}
\title{On the violation of a local form of the Lieb-Oxford bound}


\author{J. G. Vilhena}
\email[Electronic address:\;]{guilhermevilhena@gmail.com}
\affiliation{Departamento de F\'{i}sica Te\'{o}rica de la Materia
Condensada, Universidad Aut\'{o}noma de Madrid, Campus de Cantoblanco, Madrid
28049, Spain} 
\affiliation{Universit\'e de Lyon, F-69000 Lyon, France and
LPMCN, CNRS, UMR 5586, Universit\'e Lyon 1, F-69622 Villeurbanne, France}
\author{E. R{\"a}s{\"a}nen}
\email[Electronic address:\;]{erasanen@jyu.fi}
\affiliation{Nanoscience Center, Department of Physics, 
University of Jyv{\"a}skyl{\"a}, FI-40014 Jyv{\"a}skyl{\"a}, Finland}
\author{L. Lehtovaara}
\email[Electronic address:\;]{lauri.lehtovaara@gmail.com}
\affiliation{Universit\'e de Lyon, F-69000 Lyon, France and
LPMCN, CNRS, UMR 5586, Universit\'e Lyon 1, F-69622 Villeurbanne, France}
\affiliation{Nanoscience Center, Department of Physics, 
University of Jyv{\"a}skyl{\"a}, FI-40014 Jyv{\"a}skyl{\"a}, Finland}
\author{M. A. L. Marques}
\email[Electronic address:\;]{marques@tddft.org}
\affiliation{Universit\'e de Lyon, F-69000 Lyon, France and
LPMCN, CNRS, UMR 5586, Universit\'e Lyon 1, F-69622 Villeurbanne, France}


\date{\today}
\begin{abstract}
In the framework of density-functional theory, several popular 
density functionals for exchange and correlation have been constructed
to satisfy a {\em local} form of the Lieb-Oxford bound. In its
original global expression, the bound represents a rigorous lower limit 
for the indirect Coulomb interaction energy. Here we employ
exact-exchange calculations for the G2 test set to show
that the local form of the bound is violated in an extensive range of
both the dimensionless gradient and the average electron density.
Hence, the results demonstrate the severity in the usage of the local 
form of the bound in functional development. On the other hand,
our results suggest alternative ways to construct accurate 
density functionals for the exchange energy.
\end{abstract}
\maketitle


\section{Introduction}

Density-functional theory~\cite{dft} (DFT) is one of the most popular electronic
structure methods that, for a large variety of systems,  produces accurate
results with a relatively small computational cost. Functionals used for the
exchange and correlation energy $E_{xc}[n]$ and potential $V_{xc}[n]$ play a
central role in DFT~\cite{perdew_density_2003}.  In fact, DFT became a
mainstream method in Quantum Chemistry only after
significant developments beyond the local-density approximation (LDA), first and
foremost the generalized-gradient approximations (GGAs).

Most GGAs are built in a way that $E_{xc}[n]$ satisfies a set of known exact
conditions. These conditions,  together with an ansatz of the gradient form,
enable GGAs to account for inhomogeneities in the electronic density and thus
improve upon the LDA in most cases. Two of the most prominent examples of the
GGA are that of Becke from 1988~\cite{b88} (B88) and the
Perdew-Burke-Ernzerhof~\cite{perdew_generalized_1996} (PBE)  functional. In the
latter, seven known exact conditions to the $E_{xc}[n]$ are imposed to a simple
ansatz. The result is an outstandingly accurate functional that performs well
on a wide range of  systems. Currently, PBE is the most popular functional for
material applications, whereas B88 -- embedded in the B3LYP hybrid
functional~\cite{b3lyp} -- is the most popular functional in
Quantum Chemistry~\cite{burke}.

An important exact condition for $E_{xc}[n]$ is the Lieb-Oxford (LO)
bound~\cite{lieboxford}. The bound sets a rigorous lower limit for the indirect
(quantum mechanical) part of the total Coulomb interaction energy. Hence, on one
hand, the LO bound is a fundamental  condition in many-particle physics and
relates to the analysis of the stability of matter~\cite{Spruch1991}.  On the
other hand, the LO bound must be satisfied by all density functionals. In DFT
this requirement can be conveniently formulated through the expression of the
bound in terms of the LDA exchange energy (see below). 

Levy and Perdew \cite{levy_tight_1993,perdew_density_2003} have suggested that
in order to satisfy the LO bound, the {\em exchange-energy density}
(exchange-energy integrand) must also be bounded. This local condition, however,
is  stronger than the global LO bound. In spite of the ambiguity, several
functionals such as PBE rely on this local form to impose an extra constraint on
the functional. In fact, to the best of our knowledge, the LO bound has been
applied solely in its local form in the development of functionals~\cite{perdew_density_2003}.

The use of the local LO bound has been questioned in several
works~\cite{zhang,lacks}, but the severity of  the approximation, i.e., the
extent to which the local bound is violated, has not been explored until now. 
Addressing the similarities and differences between the global and local LO
bound would be of particular importance in view of  recent studies on the
tightening of the LO bound~\cite{odashimacapelle,rpcp}, and the consequent
developments or revisions of density functionals. 

In this work we use the well-known G2 test set~\cite{curtiss} to analyze to
which extent, and in which range of parameters, the local form of the LO bound is
violated.  The G2 set of 148 molecules is commonly used to gauge the accuracy
and predictive abilities of a given computational method, and it represents a
broad range of chemical environments~\cite{curtiss,grossman}.

Our analysis is carried out in the following order. In Sec.~\ref{bounds} we
review and address the differences between the  global and local forms of the LO
bound within DFT. In Sec.~\ref{results} we present our results for closed-shell
molecules in the G2 set; here we  solve exact exchange (EXX) energies within
the  Krieger-Li-Iafrate~\cite{KLI} (KLI) approximation  and demonstrate the
violation of the local bound. In the same context, we assess the local performance
of different density functionals with respect to the enhancement  factor. In
addition, we compare the relation between the violation of the local bound and
the distance from the global LO bound, and  consider the spatial dependence of
the violation of the local bound. Finally, in Sec.~\ref{outlook} we summarize
our findings and discuss how our results could be used in the development of
density functionals.


\section{Lieb-Oxford bound}\label{bounds}

\subsection{Global bound}\label{global}

The LO bound in its original, global form applies to all three-dimensional (3D)
non-relativistic, Coulomb-interacting systems. The bound can be expressed
in terms of the indirect part of the interaction energy as 
\begin{equation} 
\label{lo}
W_{xc}[\Psi]\equiv\left<\Psi|\hat{V}_{ee}|\Psi\right>-E_{H}[n]\geq-\;
C\int\nolimits d^{3}r\, n^{4/3}(\textbf{r})
\ ,
\end{equation}
where ${\hat V}_{ee}=\sum_{i>j}|{\mathbf r}_i-{\mathbf r}_j|^{-1}$ is 
the Coulombic interaction operator and $E_{H}[n]$ is the classical Hartree
energy. 
The expectation value of $\hat{V}_{ee}$ is calculated over
any normalized many-body wave function
$\Psi(\mathbf{r}_{1},...,\mathbf{r}_{N})$ with the corresponding density 
$n(\mathbf{r})$. For the prefactor $C$ 
Lieb~\cite{lieb} originally found a value $C^{{\rm L}}=8.52$, 
but this was later refined by Lieb and Oxford~\cite{lieboxford} to 
$C^{{\rm LO}}=1.68$ and numerically by Chan and Handy~\cite{ch} 
to $C^{{\rm CH}}=1.64$. Recently, using nonrigorous but physical
arguments the bound was tightened further to $C=1.44$ (Ref.~\cite{rpcp}).

Remarkably, the right-hand side of Eq. (\ref{lo}) has a 
form similar to the LDA for the the exchange energy, i.e., 
\begin{equation}
\label{eq:lda_x}
E_{x}^{{\rm LDA}}[n]=-A\int\nolimits d^{3}r\,n^{4/3}(\textbf{r})
\end{equation}
with $A=3^{4/3}\pi^{-1/3}/4$.
Moreover, the left-hand side of Eq. (\ref{lo}) can be expressed
as a density functional $W_{xc}[n]$ corresponding to the
minimization of $\left< \Psi \left| \hat{T}+ 
\hat{V}_{ee}\right| \Psi \right> $, so that the
ground-state density $n(\mathbf{r})$ is produced. Following the
definition of the exchange-correlation energy $E_{xc}$ we can now write
\begin{equation}
W_{xc}[n]\leq W_{xc}[n]+T_c[n]\equiv E_{xc}[n] \leq E_x[n],
\end{equation}
where the first inequality is justified by the fact that the kinetic-energy part
of the correlation energy is always non-negative, i.e., $T_c \geq 0$.
The second inequality follows from the non-positiveness of the correlation
energy $E_c[n]=E_{xc}[n]-E_x[n]\leq 0$. This is straightforward to see
in the constrained-search definition of the correlation functional~\cite{levy}.

Combining the above relations leads to a simple expression of the 
global bound:
\begin{equation}
E_{x}[n]\geq \lambda\, E_{x}^{{\rm LDA}}[n],
\end{equation}
or, alternatively,
\begin{equation}
\frac{E_{x}[n]}{E_{x}^{{\rm LDA}}[n]}\leq \lambda,
\end{equation}
where $\lambda=C/A$. We note that, in this definition, 
$\lambda$ is a {\em number} that in principle is universal. 
Recent studies~\cite{rpcp,paola} on 
the LO bound have focused on finding maximum values for a 
density-functional $\lambda[n]$ or function $\lambda(N)$, with an 
aim to tighten the universal value for $\lambda$. Indeed, the procedure
in Ref.~\cite{rpcp} led to a conclusion that
the bound can be tightened to $\lambda\approx 1.96$ 
corresponding to $C=1.44$. In the following,
however, we will refer to the original LO value of
$C^{{\rm LO}}=1.68$ corresponding to $\lambda=2.27$.

\subsection{Local bound}\label{local}

Before introducing the local bound it is useful to write
the exchange-correlation energy in the form of the standard
GGA ansatz, i.e.,
\begin{equation} 
\label{exc}
E_{xc}[n]=\int\nolimits d^{3}r\,n(\textbf{r})\epsilon_{xc}^{\rm 3DEG}[n]\mathcal{F}_{xc}[n,\nabla n,\ldots],
\end{equation}
where $\epsilon_{xc}^{\rm 3DEG}$ is the exchange-correlation energy per particle
in the 3D homogeneous electron gas (3DEG), and $\mathcal{F}_{xc}$ is the enhancement 
factor including the corrections to the LDA. By definition it is non-negative,
as well as its components $\mathcal{F}_{x}$ and  $\mathcal{F}_{c}$. 
In the standard GGA ansatz, $\mathcal{F}_{xc}$ is written as a functional of $n$ and $\nabla n$,
or more conveniently, in terms of the dimensionless density gradient 
$s=\left|\nabla n\right|/(2k_{F} n)$ (with $k_{F}=\left(3\pi^{2}n\right)^{1/3}$
being the Fermi momentum) and the Wigner-Seitz radius
$r_s=3^{1/3}(4\pi n)^{-1/3}$. As indicated by the symbol ``$\ldots$'' in 
Eq.~(\ref{exc}), $\mathcal{F}_{xc}$ generally depends on other quantities and
thus the expression refers to the {\em exact} $E_{xc}[n]$.

Combining the above integral form for $E_{xc}$ and the relations of the previous 
section we can rewrite the {\em global} LO bound as
\begin{equation}
\label{second}
\int\nolimits d^{3}r\,n(\textbf{r})\,\left|\epsilon_{x}^{\rm 3DEG}[n]\right|\,\mathcal{F}_{x} \leq
\lambda\int\nolimits d^{3}r\,n(\textbf{r})\,\left|\epsilon_{x}^{\rm 3DEG}[n]\right|,
\end{equation}
where we take the absolute values to deal with only non-negative quantities.
The {\em local} bound suggested by Levy and Perdew~\cite{levy,perdew_density_2003} 
(originally expressed for $\mathcal{F}_{xc}$)
is written for the integrands of Eq.~(\ref{second}), i.e.,
\begin{equation}
\label{shit}
\mathcal{F}_{x} \leq \lambda.
\end{equation}
If this local condition is satisfied, the global inequality 
in Eq.~(\ref{second}) is trivially satisfied as well. However, the 
reverse implication obviously does not hold. In other words,
Eq.~(\ref{shit}) may be violated without the violation of the 
global bound. In particular, the local bound in Eq.~(\ref{shit}) might be 
considerably too strict to be a well-reasoned condition in the 
development of density functionals. This will be explicitly demonstrated 
in the following section.


\section{Results}\label{results}

We use the {\sc octopus} code~\cite{octopus} to calculate the EXX  energies and
the corresponding EXX energy densities (per particle)  $\epsilon_x^{\rm EXX}$
for closed-shell molecules in the G2 test set. We
use the norm-conserving Hartwigsen-Goedecker-Hutter 
pseudopotentials~\cite{HGH}. The EXX results are obtained from the
optimized-effective-potential~\cite{oep} (OEP) scheme within the
KLI approximation that neglects the so-called orbital shifts in the full OEP.
Apart from  special cases such as long atomic chains~\cite{chains} the KLI
approximation has been shown to be extremely accurate with respect to the full
OEP~\cite{engel}.
 
\begin{figure}
\includegraphics[width=0.99\columnwidth]{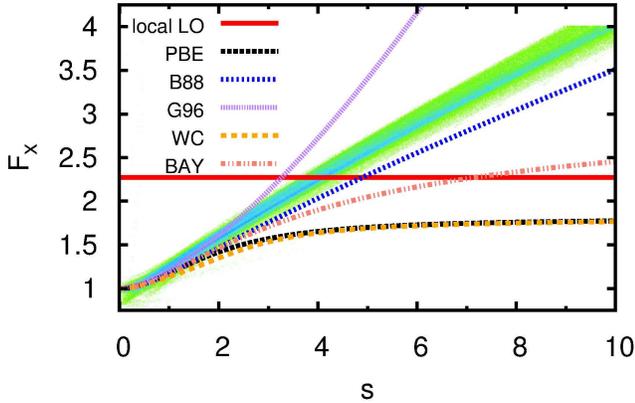}
\caption{(Color online) Count histogram of exact enhancement factors for the
electronic exchange as a function of the dimensionless gradient in the G2 test
set. The results are compared to the enhancement factors of several density
functionals. The line shows the local Lieb-Oxford bound that is violated for
$\mathcal{F}_x>2.27$. }
\label{fig1}
\end{figure}

The obtained EXX energy density per particle is directly related to the exchange
enhancement factor through  
$\epsilon_x^{\rm EXX}=\epsilon_x^{\rm 3DEG}\,\mathcal{F}_x$.
Hence, after calculating both the EXX and LDA results we can visualize the
distribution of the enhancement factors with  respect to the local bound given
in Eq.~(\ref{shit}). The procedure is the following. For every coordinate 
${\bf r}_i$, within every molecule, we collect the values 
$r_s({\bf r}_i)$, $s({\bf r}_i)$, and $\mathcal{F}_x({\bf r}_i)$, 
so that we can make a statistical count histogram of $\mathcal{F}_x$ as  a
function of $s$ or $r_s$. Figure~\ref{fig1}  visualizes the situation as a
function of $s$.  The local LO bound is shown as a horizontal red (dark gray)
line, so that all the values {\em above} the line at $\mathcal{F}_x>2.27$
violate the local bound according  to Eq.~(\ref{shit}). We find significant
violation  at $s\gtrsim 3$. This range of the dimensionless gradient, 
estimating the ratio of the density variation in the  Fermi wavelength scale,
corresponds to the tail of the  electronic density. The local dependence is
studied in more detail below.

\begin{figure}
\includegraphics[width=0.99\columnwidth]{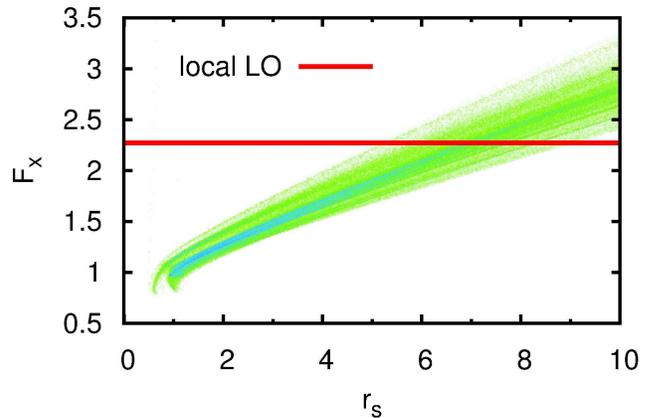}
\caption{(Color online) Same as Fig.~\ref{fig1} but as a function of
the Wigner-Seitz radius $r_s=3^{1/3}(4\pi n)^{-1/3}$.
}
\label{fig2}
\end{figure}

In Fig.~\ref{fig1} we also assess the enhancement factors of  several popular
density functionals with respect to  the EXX results and to the local bound. The
functionals include PBE (exchange only), B88, 
Gill's functional from 1996~\cite{Gill96} (G96), Wu's and Cohen's 
functional from 2006~\cite{WC} (WC), and a Bayesian fit for the enhancement
factor by Mortensen and co-workers~\cite{Bay} (BAY).
In Fig.~\ref{fig1} we can find a large variance in the behavior of
different functionals, especially at large $s$ that corresponds to the
asymptotic exponential tail of the atomic or molecular charge distribution.

As expected, PBE and WC obey the local bound {\em by definition}. However, 
they are relatively far from the histogram of exact results when 
$s\gtrsim 2$ and thus miss the correct asymptotic behavior of $\mathcal{F}_x$.
BAY has a similar trend but follows the EXX to larger values of $s$, as it
is a Pad\'e fit to experimental atomization energies of a subset of 
the G2 test set. 

Interestingly, B88 performs best of all the tested approximations. 
This is due to the fact that B88 has a parameter chosen to 
reproduce the Hartree-Fock exchange energies for atomic systems that are 
conceptually similar to the G2 test set considered here. B88 also
by construction obeys the correct asymptotic behavior of the exchange 
energy per particle, i.e., $\epsilon_{x}(r) \rightarrow -1/(2r)$ as 
$r\rightarrow\infty$. G96 does not obey this limit and strongly deviates
from the EXX results at large $s$.

Figure~\ref{fig2} shows the distribution of the EXX enhancement factors as a
function of the Wigner-Seitz radius $r_s$. The local bound is violated at
$r_s\gtrsim 5$ corresponding to densities around $n\lesssim 0.002$\,bohr$^{-1}$. Considering
a local version of a recently proposed tighter global bound~\cite{rpcp} 
with $\lambda=1.96$ (instead of the LO one) would lead to violation already at 
$r_s\sim 4$ ($n\sim 0.004$\,bohr$^{-1}$). In any case, the violation clearly seems to occur
in the tail of the electronic density as shown explicitly below.

\begin{figure}
\includegraphics[width=0.99\columnwidth]{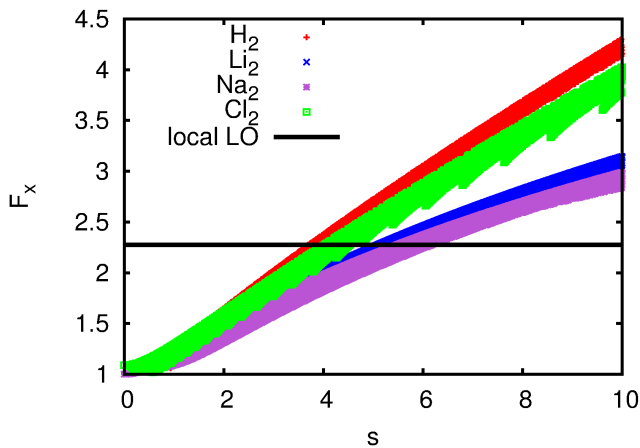}
\caption{(Color online) Enhancement factor as a function of the 
dimensionless gradient $s$ for four diatomic molecules.}
\label{fig3}
\end{figure}

In Fig.~\ref{fig3} we have a closer look at the enhancement  factor as a
function of $s$ for four diatomic molecules  including H$_2$, Cl$_2$, Na$_2$,
and Li$_2$. H$_2$ and Cl$_2$ violate the local LO bound more 
rapidly than Na$_2$, and Li$_2$. It is interesting to compare these 
results with the {\em distance from the  global bound} 
considered by Odashima and Capelle~\cite{mariana}.
In their study H$_2$ was found to be closest to
the global bound with $\lambda[n]\approx 1.25$.
Therefore the high tendency of H$_2$ to violate
the local bound in Fig.~\ref{fig3} is plausible
and unlikely to be purely accidental, although obviously 
there is no rigorous implication from the global to 
local bound as discussed in Sec.~\ref{bounds}.
In fact, in Ref.~\cite{mariana} Li$_2$ 
and Cl$_2 $ were found to have 
$\lambda[n]\approx 1.21$ and $1.11$, respectively, 
whereas the local bound-breaking tendency is
much higher in the latter system (Fig.~\ref{fig3}).
Therefore, the spatial distribution of 
$\mathcal{F}_x$ has an important role, so that
there is no clear correlation in the
sensitivity of a particular molecule to the global 
and local bounds. As discussed above, the only rigorous implication is 
the fact that if the local bound is obeyed (as in PBE), 
the global bound is obeyed as well.

\begin{figure}
\includegraphics[width=0.99\columnwidth]{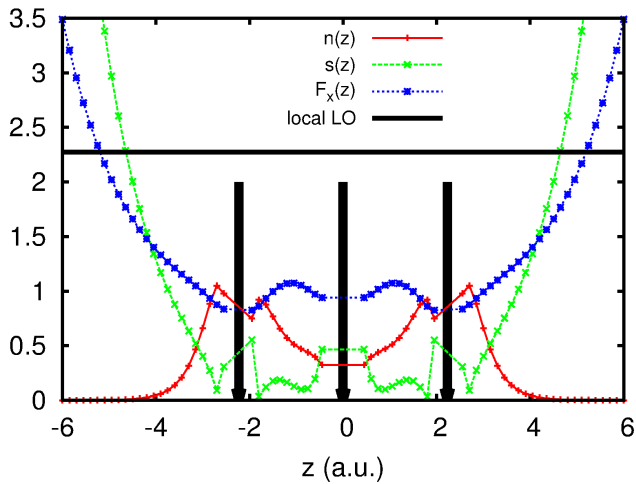}
\caption{(Color online) Spatial behavior of the enhancement
factor across the molecular axis of CO$_2$.
The density $n$ and dimensionless gradient $s$
are also shown (in arbitrary units).
The arrows mark the position of the atoms.}
\label{fig4}
\end{figure}

Finally we give an example of the spatial dependence
of the enhancement factor. Figure~\ref{fig4}
shows $\mathcal{F}_x$ along the molecular axis
of CO$_2$, plotted together with the density and
the dimensionless gradient (in arbitrary units). 
The arrows mark the position of the atoms. 
Furthermore, for clarity we removed the points inside the
cores of the pseudopotentials. The figure confirms
our statement within Fig.~\ref{fig1} that the local bound
is violated in the tail of the electronic density
where $n$ is small and $s$ is large. Similar
spatial behavior was found in all the cases
that we checked in this detail.

\section{Conclusions and outlook}\label{outlook}

We have explicitly studied the conceptual difference between the
universal Lieb-Oxford bound and its local interpretation used in the
development of density functionals. Our molecular examples in the G2
test set demonstrate that the local bound is broadly violated when
exact exchange enhancement factors are considered. Therefore the use
of the local bound in the development of functionals , e.g., in
generalized-gradient approximations is questionable, even if the
condition is straightforward to implement amd the fulfillment of the
global Lieb-Oxford bound is guaranteed. In short, the local bound is
simply much stricter that has been previously thought.

Our exact-exchange results pinpoint the violation  of the local bound to
$s\gtrsim 3$ or  $r_s\gtrsim 5$ corresponding to the tail of the electronic
density. We find no clear  correlation between the distance from the global
bound and the degree of violation of the local bound; this demonstrates the 
complexity of the enhancement factor in the tail region throughout the 
ensemble of molecules.

The surprisingly uniform distribution of the exact enhancement factors in
Figs.~\ref{fig1} and \ref{fig2} suggest a construction of a density functional
according to the observed $\mathcal{F}_x(s)$. This could  be done with a fitting
procedure or, better, by tailoring a physically motivated ansatz in the
(meta-) generalized-gradient fashion that is able to reproduce the exact
$\mathcal{F}_x(s)$ to a reasonable extent. Such an ability is of particular 
importance when describing physical properties that depend on the correct
description of the tail of the electronic density, e.g., ionization and
Rydberg excitations etc. Within the construction, however, a
simultaneous implementation of the {\em global} Lieb-Oxford bound  is a tedious
(if not practically impossible) task. These aspects will be studied in detail
in our future works.

\begin{acknowledgments}
J.G.V. acknowledges support from the FCT Grant 
No. SFRH/BD/38340/2007, M.A.L.M. from the French ANR 
(ANR-08-CEXC8-008-01), and E.R. and L.L. from the 
Academy of Finland.
\end{acknowledgments}

\end{document}